\documentclass[prb,twocolumn,showpacs,preprintnumbers,amsmath,amssymb]{revtex4}

\usepackage{graphicx}
\usepackage{amsmath}
\usepackage{amsbsy}
\usepackage{dcolumn}
\usepackage{bm}

\begin{document}

\title{Novel Josephson Effects in $d$-wave Superconductor Junctions with Magnetic Interlayers}

\author{Brian M. Andersen$^1$, Yu. S. Barash$^2$, S. Graser$^3$, and P. J. Hirschfeld$^3$}

\affiliation{
$^1$Nano-Science Center, Niels Bohr Institute, University of Copenhagen, Universitetsparken 5, DK-2100 Copenhagen, Denmark\\
$^2$Institute of Solid State Physics, Russian Academy of Sciences,
Chernogolovka, Moscow reg., 142432 Russia\\
$^3$Department of Physics, University of Florida, Gainesville,
Florida 32611-8440, USA}

\date{\today}

\begin{abstract}
We calculate the dc supercurrent through a Josephson tunnel junction
consisting of an antiferromagnetic (AF) or ferromagnetic (F)
interlayer sandwiched between two $d$-wave superconductors ($d$).
Such junctions exhibit a rich dependence of the Josephson current on
the interlayer parameters, including the possibility of $0-\pi$
transitions with varying temperature or interlayer thickness.
Furthermore, we study $d$/I/$d$ junctions when the dSC leads include
subdominant magnetic correlations. Induced magnetism near the
interface can strongly diminish the critical current for 110
oriented junctions whereas no suppression exists for the 100
orientation. This may help resolve a long-standing puzzle of the
critical current versus grain boundary angle in high-$T_c$
superconductors.
\end{abstract}

\pacs{74.45.+c, 74.50.+r, 75.50.Ee, 74.72.-h} \maketitle

\section{Introduction}
Interfaces and Josephson junctions between superconductors and
magnetic materials can generate novel low-energy spin-dependent
Andreev bound states leading to highly unconventional quantum
transport properties. For instance, junctions consisting of $s$-wave
superconductors ($s$) and ferromagnetic (F) metals have attracted
great interest in recent years\cite{buzdin05}. Such junctions have
been shown to exhibit so-called $0-\pi$
transitions\cite{ryazanov01,aprili02}, where, depending on the
temperature $T$ and the width $L$ of the interlayer, the ground
state is characterized by an internal phase difference of $\pi$
between the two superconductors. This effective negative Josephson
coupling is similar to what can happen when tunnelling through
magnetic impurities\cite{bulaev}. The possibility of $0-\pi$
transitions may be utilized as a basis for future quantum
qubits\cite{beasley}, constituting an important example in the field
of superconducting spintronics\cite{sarma04}.

Another promising situation involves interfaces between
antiferromagnets and superconductors. In this case spin dependent
quasiparticle reflection at the AF surface, so-called Q-reflection,
combined with Andreev reflection on the superconducting side, can
lead to new low-energy bound states with important consequences for
the associated proximity effect\cite{bobkova05,andersen05}.
Furthermore, in $s$/AF/$s$ Josephson junctions these bound states
can enhance the critical current $J_c$ at low $T$\cite{bobkova05},
and lead to $0$- or $\pi$-junction behavior depending on $T$ and
thickness of the AF interlayer\cite{andersen06}. For $s$/AF/$s$
junctions the $0-\pi$ behavior is a true even-odd effect arising
from qualitatively different spectra of the Andreev bound states
caused by different symmetries of the odd and even AF
interfaces\cite{andersen06}.

We study the Josephson current through in-plane $d$/AF/$d$ tunnel
junctions. Such junctions have not been studied before
theoretically. Interestingly, our results are also relevant for
d/F/d junctions. Based on both analytical calculations and numerical
solutions of the Bogoliubov-de Gennes (BdG) equations, we determine
the criteria for $0-\pi$-junction behavior and predict unusual $T$
dependence of the critical current $J_c(T)$.

Intrinsic $d$/AF/$d$ junctions may already be present in the case of
high-T$_c$ grain boundaries (GB) which induce AF surface states.
Below, we also study the critical current through GB by modeling
them as $d$/I/$d$ junctions, where I is an insulating layer but
where the leads contain subdominant magnetic correlations which
become important near order parameter-suppressing interfaces. Both
kinds of junctions mentioned above are cousins of the unconventional
$d$/I/$d$ junctions with uncorrelated leads which exhibit an unusual
$1/T$ behavior of $J_c(T)$ at low $T$ as well as possible (depends
on misorientation angle) $T$-induced $0-\pi$
transitions\cite{bbr96,tanakakas96}. The experimental observation of
these effects is notoriously difficult due to the complexity of the
barrier interface, characterized, in particular, by facetting, twins
and especially by many high transmission channels. Only recently
have the main features associated with mid-gap state contribution to
the Josephson current been observed in experiments
\cite{ilichev01,blamire04,blamire05}.

\section{Model}
The Hamiltonian is defined on a two-dimensional (2D)
square lattice (lattice constant $a=1$)
\begin{eqnarray}\label{hamiltonian}
\hat{H}= &-& t \sum_{\langle ij \rangle\sigma}
\hat{c}_{i\sigma}^{\dagger}\hat{c}_{j\sigma} + \sum_{\langle ij
\rangle} \left( \Delta_{ij}
\hat{c}_{i\uparrow}^{\dagger}\hat{c}_{j\downarrow}^{\dagger} +
\mbox{H.c.} \right) \nonumber\\ &-& \sum_{i\sigma} \mu
\hat{n}_{i\sigma} + \sum_{i} m_i \left(\hat{n}_{i\uparrow} -
\hat{n}_{i\downarrow} \right).
\end{eqnarray}
Here, $\hat{c}_{i\sigma}^{\dagger}$ creates an electron of spin
$\sigma$ on the site $i$, $t$ is the hopping matrix element, $\mu$
is the chemical potential, and $\Delta_{ij}$ and $m_i$ denote the
superconducting and magnetic order parameters, respectively. The
associated BdG equations are given by
\begin{equation}\label{BdG}
\sum_j \left( \begin{array}{cc} {\mathcal{K}}^{+}_{ij,\sigma}&
{\mathcal{D}}_{ij,\sigma} \\
{\mathcal{D}}^*_{ij,\sigma} & -{\mathcal{K}}^{-}_{ij,\sigma}
\end{array}\!\right)\! \left( \begin{array}{c} u_{n\sigma}(j) \\
v_{n\overline{\sigma}}(j) \end{array}\!\right)\! =\!
E_{n\sigma}\! \left( \begin{array}{c} u_{n\sigma}(i) \\
v_{n\overline\sigma}(i) \end{array}\! \right),
\end{equation}
where ${\mathcal{K}}^{\pm}_{ij}=-t\delta_{\langle
ij\rangle}+(\pm\sigma m_i-\mu)\delta_{ij}$, with $\sigma=+1/-1$ for
up/down spin and $\delta_{ij}$ and $\delta_{\langle ij \rangle}$ are
the Kronecker delta symbols connecting on-site and nearest neighbor
sites, respectively. The net magnetization is $M_i=m_i/U=\frac{1}{2}
\left[ \langle \hat{n}_{i\uparrow} \rangle\ - \langle
\hat{n}_{i\downarrow} \rangle \right]$, and the off-diagonal block
${\mathcal{D}}_{ij}$ describes $d$-wave pairing
${\mathcal{D}}_{ij}=-\Delta^d_{ij}\delta_{\langle ij \rangle}$,
where $\Delta^d_{ij}=-V\langle \hat{c}_{i\uparrow}
\hat{c}_{j\downarrow} - \hat{c}_{i\downarrow}
\hat{c}_{j\uparrow}\rangle/2$. The coupling constants $U$ ($V$) are
non-zero on (off) the $L$ atomic chains constituting the AF
interlayer. By Fourier transform parallel to the interface, we
obtain an effective 1D problem at the expense of introducing an
additional parameter $k_y$. The dc Josephson current $j_{rr'}$
between two neighboring sites $r$ and $r'$ is obtained from
$j_{rr'}=-(iet/\hbar) \sum_\sigma \left[ \langle
\hat{c}_{r\sigma}^{\dagger}\hat{c}_{r'\sigma} \rangle - \langle
\hat{c}_{r'\sigma}^{\dagger}\hat{c}_{r\sigma} \rangle \right]$. For
more details on the numerical and analytical approaches, we refer
the reader to Refs. \onlinecite{bobkova05}-\onlinecite{andersen06}.

\section{Results}
\begin{figure}[t]
\begin{minipage}{.49\columnwidth}
\includegraphics[clip=true,width=.98\columnwidth]{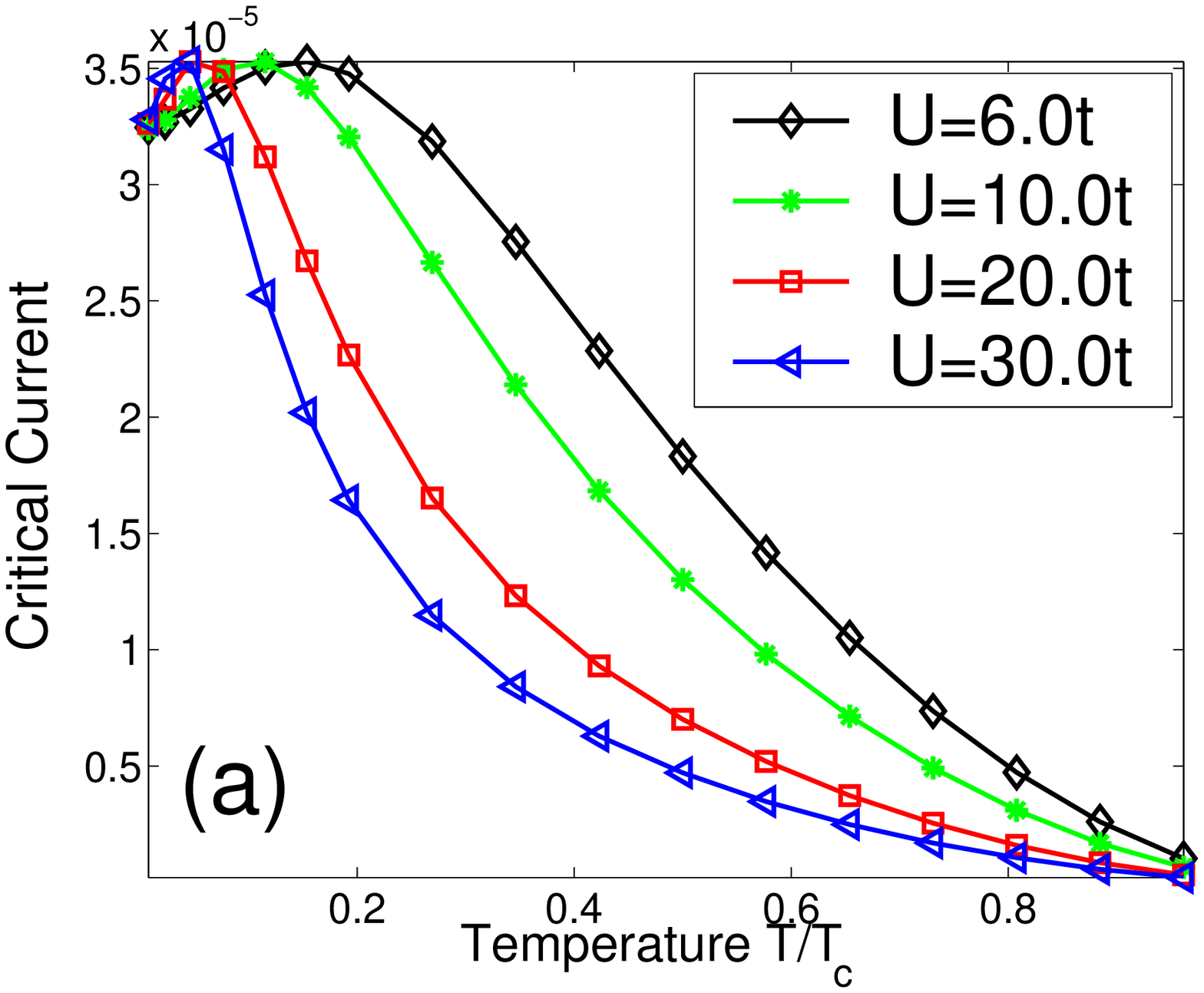}
\end{minipage}
\begin{minipage}{.49\columnwidth}
\includegraphics[clip=true,width=.98\columnwidth]{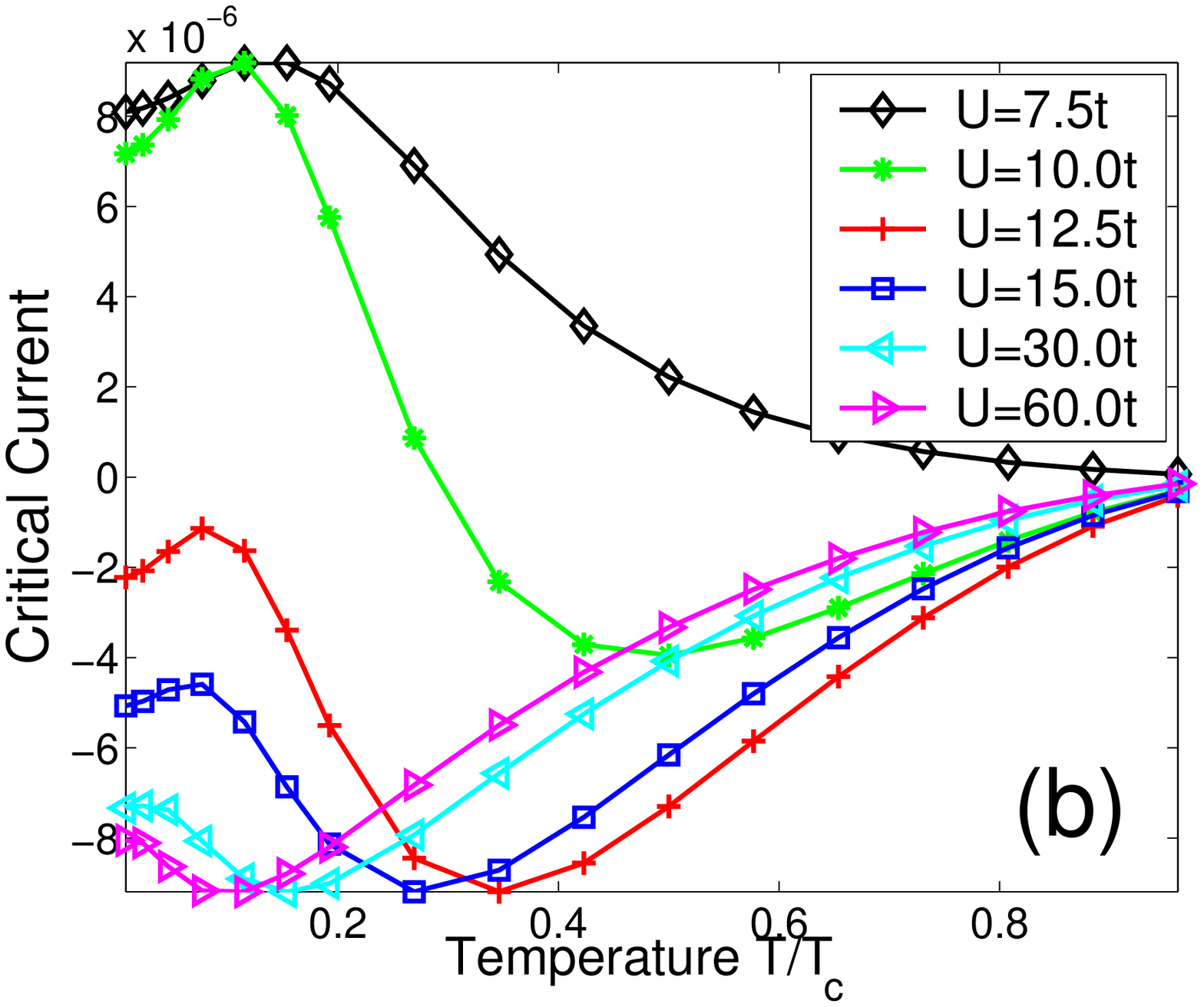}
\end{minipage}\\
\begin{minipage}{.49\columnwidth}
\includegraphics[clip=true,width=.98\columnwidth]{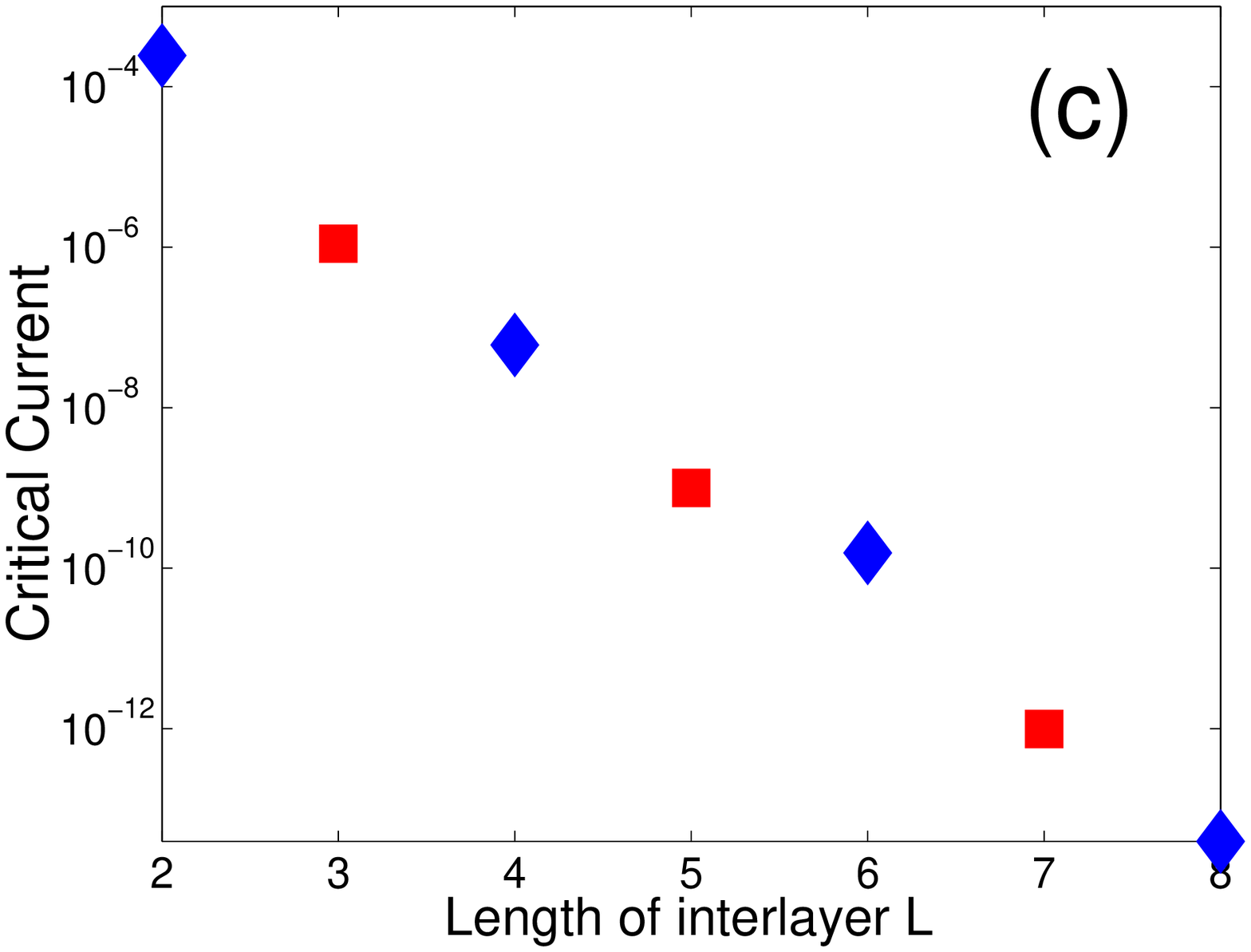}
\end{minipage}
\begin{minipage}{.49\columnwidth}
\includegraphics[clip=true,width=.98\columnwidth]{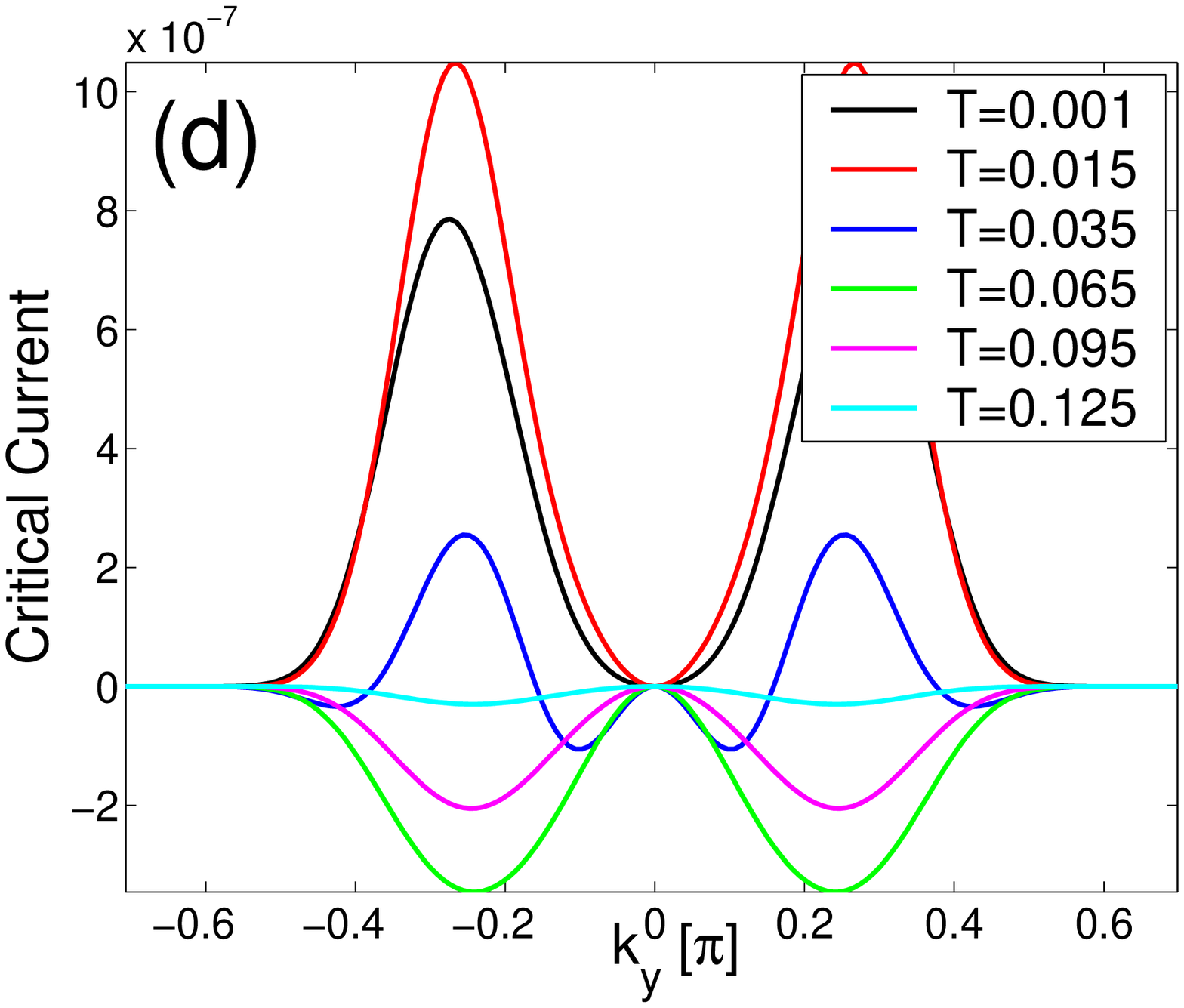}
\end{minipage}\\
\caption{(Color online) $J_c(T)$ vs $T$ for a range of $U$ for (a)
$L=6$ and (b) $L=5$. For clarity, the curves have been normalized to
give the same absolute value at the maximum as the curve with the
smallest U. (c) $J_c$ at $T=0$ as a function of interlayer thickness
$L$. The red squares correspond to negative $J_c$ and have been
multiplied by $-1$ in this semi-log plot. (d) the $k_y$-resolved
current shown for $L=5$ and $U=10t$ (along green curve in (b)). For
all results shown here: $\mu=0$ and $V=t$. The latter leads to
$\xi/L \sim 1-2$ in agreement with the short in-plane coherence
length of cuprate superconductors.} \label{fig:1}
\end{figure}

For $s$/AF/$s$ junctions, the $0-\pi$ behavior as a
function of interlayer thickness $L$ exists both for 100 and 110
orientations\cite{andersen06}. This is not the case for $d$/AF/$d$
junctions, where the 100 case displays only 0-junction
characteristics with an Ambegaokar-Baratoff like dependence of
$J_c(T)$. Therefore, we focus on the more interesting 110 oriented
$d$/AF/$d$ junctions. We discuss only identical (and identically
oriented) junctions, and restrict ourselves to the tunnelling limit
where the current-phase relation is sinusoidal, and $J_c=J(\pi/2)$.
The 110 oriented $d$/AF/$d$ junctions are categorized further into
d/AFeven/d and d/AFodd/d junctions depending on whether the
interlayer consists of an even or odd number of chains,
respectively. In Fig. \ref{fig:1}(a) and \ref{fig:1}(b) we show
typical self-consistent results for $J_c$ as a function of $T$ for
even and odd interlayer-chains, respectively. As seen from Fig.
\ref{fig:1}(a), d/AFeven/d are 0-junctions with a $1/T$-like
dependence of $J_c$ in the large-$U$ limit. The small dip in $J_c$
at low $T$ is caused by the finite width of the interlayer and
disappears in the limits $\xi/L,U\rightarrow \infty$. As shown in
Fig. \ref{fig:1}(b), $J_c(T)$ in 110 d/AFodd/d junctions exhibits a
surprisingly rich $T$ dependence: as $U$ is increased, the pure
0-junction at low $U$ becomes a $\pi$-junction at high $T$, crossing
over to 0-junction behavior at some $T^*$ which eventually vanishes
in the large-$U$ limit where $J_c(T) \sim -1/T$. The systematic
$0-\pi$-junction oscillations versus interlayer thickness $L$ is
shown in Fig. \ref{fig:1}(c). The $k_y$-resolved current
corresponding to parameters similar to the green curve in Fig.
\ref{fig:1}(b), is shown in Fig. \ref{fig:1}(d). The same momentum
region contributes to the current at all $T$, a fact which will ease
the analytical interpretation presented in the next section. Results
qualitatively similar to those shown in Fig. \ref{fig:1} can be also
obtained for thicker junctions with smaller values of $U/t$.

We now study $d$/AF/$d$ junctions within a quasiclassical approach
where, as usual, all characteristic energies are assumed to be much
less than the Fermi energy $E_F$. We assume for the coherence length
$\xi \gg a,L$, and the junction properties are conveniently
expressed in terms of the scattering $\cal{S}$ matrix containing the
reflection and transmission amplitudes. The Josephson current is
carried entirely by phase-dependent Andreev bound states.

For the 100 orientation the $d$-wave order parameter does not change
its sign in specular reflection, but it changes sign in Q-reflection
processes. By contrast, in the 110 case the $d$-wave order parameter
changes its sign both in specular and in Q-reflection events. An
important manifestation of this physical difference between effects
of Q-reflection for different AF-interface orientations is that the
$0-\pi$ transition does not take place for 100 orientation in
$d$-wave junctions, but the transition is, in general, present in
110 $d$-wave, as well as 100 $s$-wave junctions. More formally, in
the 110 case the specular and Q-reflection possess identical
outgoing group velocities and form the outgoing flow along one and
the same direction. This permits the reduction of the problem to a
standard situation with conventional number of incoming and outgoing
waves, which determines the rank of the S-matrix. This is not the
case for the 100 orientation, when specular and Q reflection should
be considered separately. This increases the rank of the S-matrix
and makes ultimate results for 100 junctions with finite
transparencies strongly different compared to the 110 case. In the
following we focus solely on the 110 orientated interfaces.

For d/AFodd/d junctions, the general structure of the $\cal S$
matrix is similar to that of d/F/d junctions with symmetric F
interfaces. This follows from the fact that in the (110)
orientation and for an odd number of chains in the interlayer, all
spins are aligned in the outermost chains. For (110) d/AFeven/d
junctions, the outermost chains of the AF interface have opposite
spin polarizations (but still all aligned within each chain) and
the $\cal S$ matrix is isomorphic to the three-layer FIF interface
with antiparallel orientations of the two F layers\cite{bbk02}.
The existence of the even-odd ($0-\pi$) behavior shown in Fig.
\ref{fig:1}(c) follows directly from this link between the $\cal
S$ matrices for 110 $d$/AF/$d$ and d/F/d
junctions\cite{andersen06}. However, in order to understand the
$T$ dependence of $J_c(T)$ and obtain quantitative criteria for
transitions between 0- or $\pi$-junction behavior, we turn now to
the explicit calculations.

\begin{figure}[t]
\begin{minipage}{.49\columnwidth}
\includegraphics[clip=true,width=.98\columnwidth]{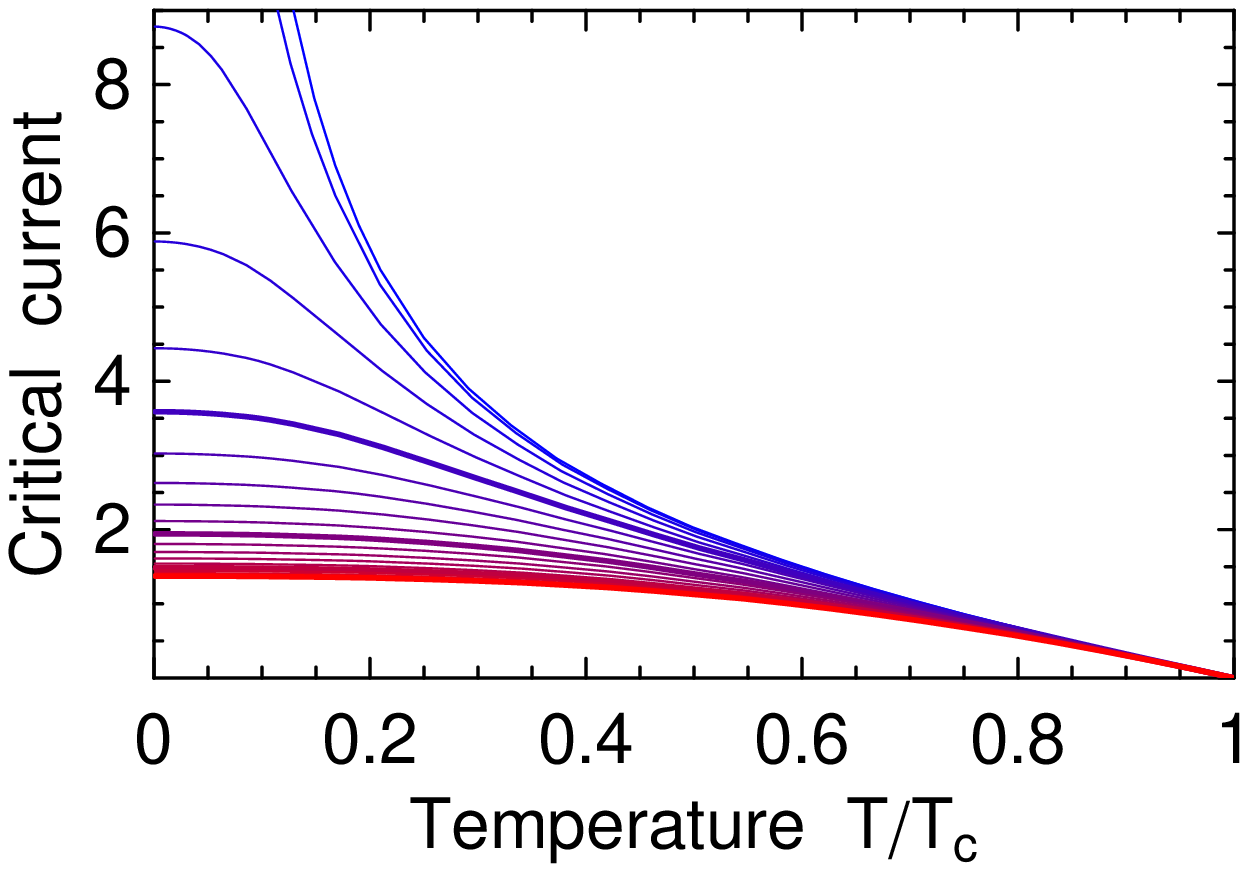}
\end{minipage}
\begin{minipage}{.49\columnwidth}
\includegraphics[clip=true,width=.98\columnwidth]{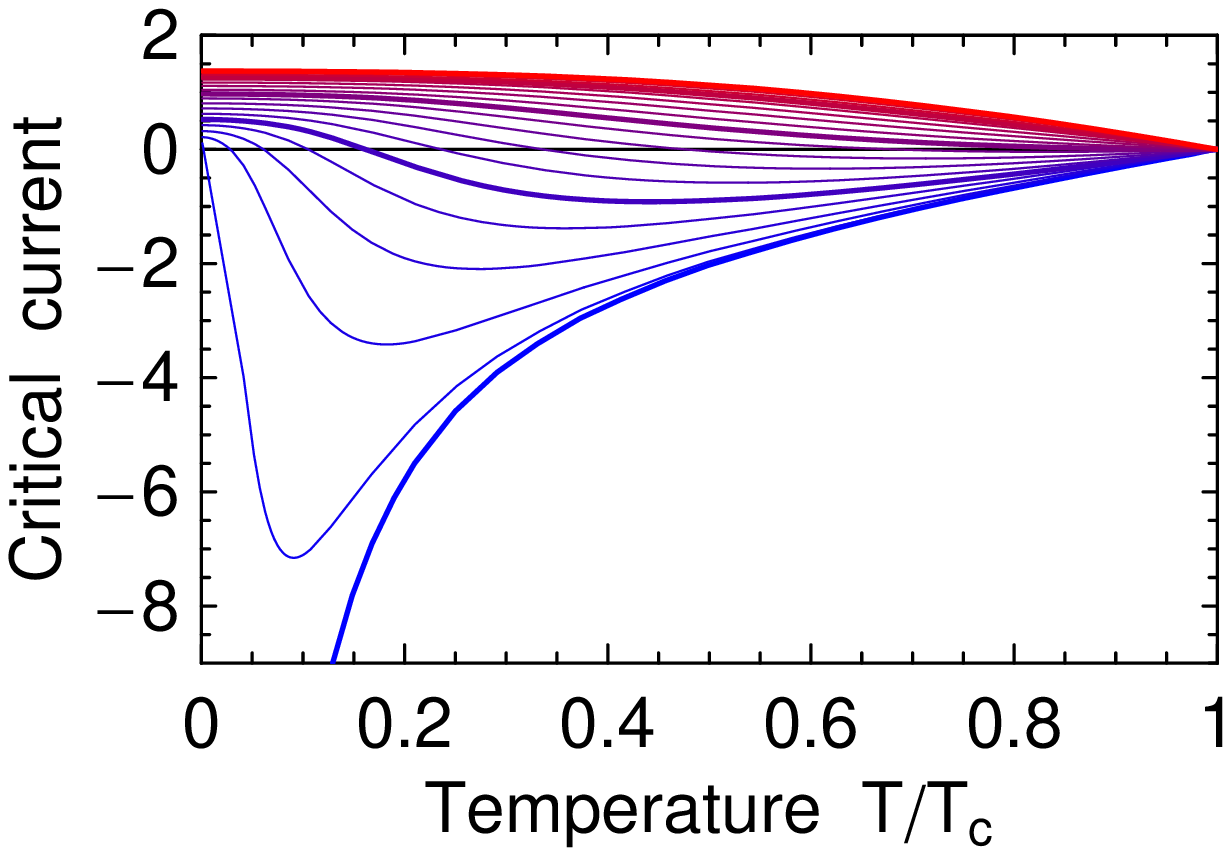}
\end{minipage}
\caption{(Color online) $J_c(T)$ obtained from quasiclassical
calculations in the tunnelling limit  $D=0.001$ for d/AFeven/d
(left) and d/AFodd/d (right), plotted for different spin-mixing
angles (from $\Theta=0$ (blue) to $\Theta=\pi$ (red) in intervals of
$0.05\pi$). The $k_y$ dependence of $\Theta$ has been neglected
here.} \label{fig:2}
\end{figure}

Consider first (110) d/AFeven/d junctions where the transparency
coefficients satisfy $D_\sigma=D_{\bar\sigma}=D=1-R$, resulting in
the following Josephson current
\begin{equation}
J(\chi,T)=\frac{e|\Delta^d| D\sin\chi}{\gamma} {\rm tanh}\!\left[
\frac{|\Delta^d|\gamma}{2 T}\right], \label{josmagdFIF}
\end{equation}
where $\gamma=\left(R\sin^2\frac{\Theta}{2}+
D\cos^2\frac{\chi}{2}\right)^{1/2}$, and $\chi$ is the phase
difference across the junction. Here, not only $\Delta^d$ and $D$,
but also the spin-mixing parameter $\Theta$ ($\sin\Theta(k_y
)=\left[m/2t\cos\left(k_y/\sqrt{2}\right)\right]\left\{1+\left[
m/4t\cos\left(k_y/\sqrt{2}\right)\right]^2\right\}^{-1}$), are all
$k_y$-dependent, and the total current is a sum of
Eq.\eqref{josmagdFIF} over all values of $k_y$\cite{andersen06}.
However, as seen from Fig. \ref{fig:1}(d), the $k_y$-sum is
unimportant for understanding the qualitative behavior. Eq.
\eqref{josmagdFIF} is valid for arbitrary transparency, and the
critical current $J_c(T)$ is plotted in Fig. \ref{fig:2}(left) for a
range of $\Theta$. In agreement with Fig. \ref{fig:1}(a), the
junction is always a 0-junction. Near $T_c$, Eq. \eqref{josmagdFIF}
reduces to $J(\chi,T)=e|\Delta^d|^2 D\sin\chi/2T$ which coincides
with the result in nonmagnetic (110) symmetric $d$/I/$d$ junctions.
However, at low $T$ the current (\ref{josmagdFIF}) is given by
$J(\chi,T)={e|\Delta^d| D\sin\chi}/{\gamma}$ which, in the
tunnelling limit, reduces to $J(\chi,T)={e|\Delta^d|
D\sin\chi}/{\left|\sin\frac{\Theta}{2}\right|}$. Therefore, due to
the factor $\left|\sin\frac{\Theta}{2}\right|^{-1}$, we find the
remarkable result that the current substantially exceeds the
critical current in nonmagnetic (110) $d$/I/$d$ junctions with the
same transparency coefficient $D$ (Fig. \ref{fig:2}).

Next we discuss the $T$ dependence of $J_c(T)$ for 110 d/AFodd/d
junctions. As argued above, this junction is similar to 110 d/F/d
junctions with symmetric F interlayer. In the
tunnelling limit we obtain the following expression for $J_c(T)$
($\alpha \equiv \pm 1$)
\begin{eqnarray}\nonumber
J_c(T)&=&\alpha e |\Delta^d|\sqrt{D_\sigma D_{\bar\sigma}} \left[
\sin\left(\frac{\Theta}{2}\right)\tanh\left(\frac{
|\Delta^d|\sin\left(\frac{\Theta}{2}\right)}{2 T}\right)\right.
\\
&-&\left. \frac{|\Delta^d|\cos^2\left(\frac{\Theta}{2}\right)}{2
T}\cosh^{-2}
\left(\frac{|\Delta^d|\sin\left(\frac{\Theta}{2}\right)}{2 T}\right)
\right], \label{josmagtund}
\end{eqnarray}
which is plotted in Fig. \ref{fig:2}(right) for $\alpha=1$ (see
below). The result for an arbitrary transparency can be obtained
along similar lines to Ref. \onlinecite{bb02}. In the absence of
magnetism, when $\Theta=0$, $\alpha=-1$, and for zero transparency
$D\rightarrow 0$, there are zero-energy Andreev bound states at both
$d$/I surfaces of the (110) $d$/I/$d$ junction. With increasing
$\Theta$, the mid-gap states on each $d$/I surface evolve into
spin-split Andreev states on a $d$/FI surface. For a given $k_y$,
the energies of these spin-split states are
$\varepsilon_d=\pm\Delta^d\sin\left(\frac{\Theta}{2}\right)
\label{surfstatesd}$. This is different from the $s$-wave case where
$\varepsilon_s=\pm\Delta^s\cos\left(\frac{\Theta}{2}\right)$\cite{fogel00},
and therefore the behavior of the Josephson current in s/F/s tunnel
junctions\cite{bb02} strongly differs from $d$-wave magnetic
junctions. Eq.\eqref{josmagtund} can be qualitatively understood as
follows: in tunnel junctions the surface states $\varepsilon_d$
further split and become phase-dependent due to a finite
transparency. As a result, four current-carrying interface Andreev
states exist for a given $k_y$. Eq. (\ref{josmagtund}) represents
the Josephson current carried by these states in the tunnelling
limit, when two spin-split surface states on each side of the
junction only slightly overlap through the interlayer.

In the limit of a nonmagnetic interlayer ($\Theta=0$, $\alpha=-1$),
only the second term in Eq.(\ref{josmagtund}) survives and one
obtains $J_c(T)=e |\Delta^d|^2 D/2T\label{josnemagtund}$, with the
well-known $1/T$-behavior for $d$/I/$d$ junctions. This result is
the tunnelling limit of the more general current-phase
relation\cite{riedel,tanakakas2000}
\begin{equation}
J(\chi,T)=2e\left|\Delta^d\right|
\sqrt{D}\sin\frac{\chi}{2}\tanh\left[\frac{\left|\Delta^d\right|}{2
T}\sqrt{D}\cos\frac{\chi}{2}\right] . \label{josnemagsymdId}
\end{equation}
Hence, there are no $0-\pi$ transitions in (110) $d$-wave
nonmagnetic junctions. This, however, is not the case, in the
presence of magnetic interlayers with finite spin-mixing $\Theta$.
Finite values of $\Theta$ result in the appearance of the additional
(first) term in Eq.(\ref{josmagtund}), which is comparatively small
for small $\Theta$, and for $\theta<\pi/2$ has the opposite sign
compared to the second term. The second term in
Eq.\eqref{josmagtund} is, in its turn, strongly modified due to
finite $\Theta$ at sufficiently low $T$. Indeed, it becomes
exponentially small, if $T$ is much less than the spin-split finite
energies of the Andreev states $\varepsilon_d$. At the same time,
the second term in Eq.\eqref{josmagtund} dominates the current at
higher $T$, for example, near $T_c$. For this reason the $0-\pi$
transition arises in magnetic 110 d/AFodd/d tunnel junctions under
the condition $\Theta<\pi/2$, as a result of the interplay of the
two terms with opposite signs in Eq.(\ref{josmagtund}). In
principle, the change of sign of the total current in
(\ref{josmagtund}) takes place with varying $T$ for any small value
of $\Theta$, but effects of finite transparency confine the
conditions for the presence of a $0-\pi$ transition to not too small
values of $\Theta$.

For deriving the conditions for the presence of the $0-\pi$
transition in the tunnelling limit, it is convenient to consider
two limiting cases of Eq.(\ref{josmagtund}), one near $T_c$ and
another at low $T$. Under the condition
$\frac{|\Delta^d|}{2}\sin\left(\frac{\Theta}{2}\right)\ll  T \le
T_c$, Eq.\eqref{josmagtund} reduces to the simple expression
\begin{equation}
J_c(T)=\frac{-\alpha e |\Delta^d|^2\sqrt{D_\sigma
D_{\bar\sigma}}}{2T} \cos\Theta\enspace, \label{josmagtundtc}
\end{equation}
which is suppressed by the factor $\cos\Theta$ compared to the
corresponding nonmagnetic $d$/I/$d$ junction. Eq.
\eqref{josmagtundtc} is valid, in particular, near $T_c$. Under
the opposite condition
$T\ll\frac{|\Delta^d|}{2}\sin\left(\frac{\Theta}{2}\right)$,
Eq.\eqref{josmagtund} becomes
\begin{equation}
J_c(T)=\alpha e |\Delta^d|\sqrt{D_\sigma D_{\bar\sigma}}
\left|\sin\left(\frac{\Theta}{2}\right)\right|\enspace,
\label{josmagtundlowT}
\end{equation}
which is suppressed by the factor
$\left|\sin\left(\frac{\Theta}{2}\right)\right|$ compared to
nonmagnetic $d$/I/$d$ junction. Comparing  signs of Eqs.
\eqref{josmagtundtc} and \eqref{josmagtundlowT}, it is evident that
the $0-\pi$ transition takes place with varying $T$ when
$\cos\Theta>0$, that is for $\Theta<\frac{\pi}{2}$. For $\alpha=1$
(which is the case for d/AFodd/d junctions) and
$\Theta<\frac{\pi}{2}$ the $0$-state is the ground state of the
junction, 
whereas the $\pi$-state exists near $T_c$ in qualitative agreement
with Fig. \ref{fig:1}(b). Note that $0-\pi$ transitions in s/F/s
junctions happens when the opposite inequality
$\Theta>\frac{\pi}{2}$ is satisfied\cite{bb02}. We stress that our
results \eqref{josmagtund}, \eqref{josmagtundtc} and
\eqref{josmagtundlowT} describe also the current in d/F/d junctions.

\begin{figure}[b]
\begin{minipage}{.49\columnwidth}
\includegraphics[clip=true,width=.98\columnwidth]{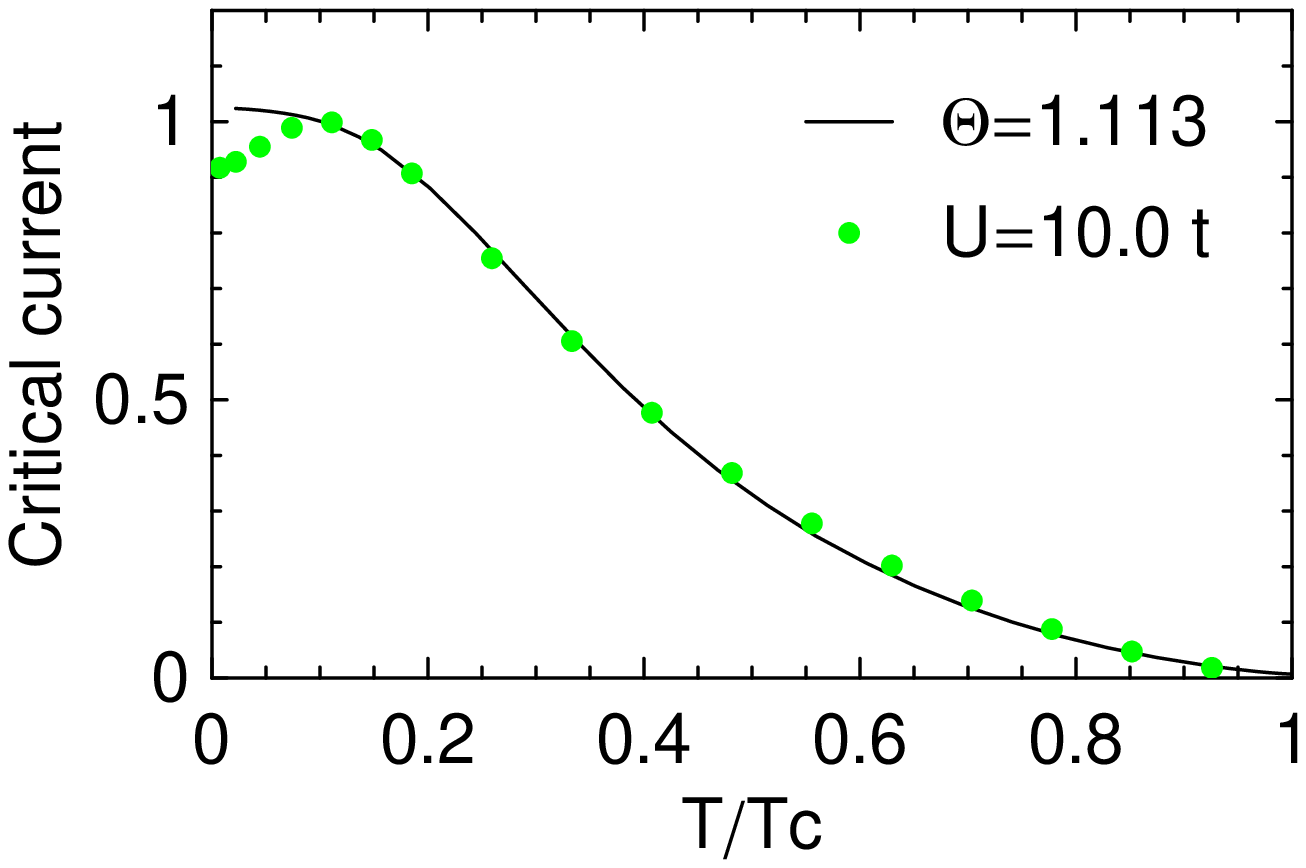}
\end{minipage}
\begin{minipage}{.49\columnwidth}
\includegraphics[clip=true,width=.98\columnwidth]{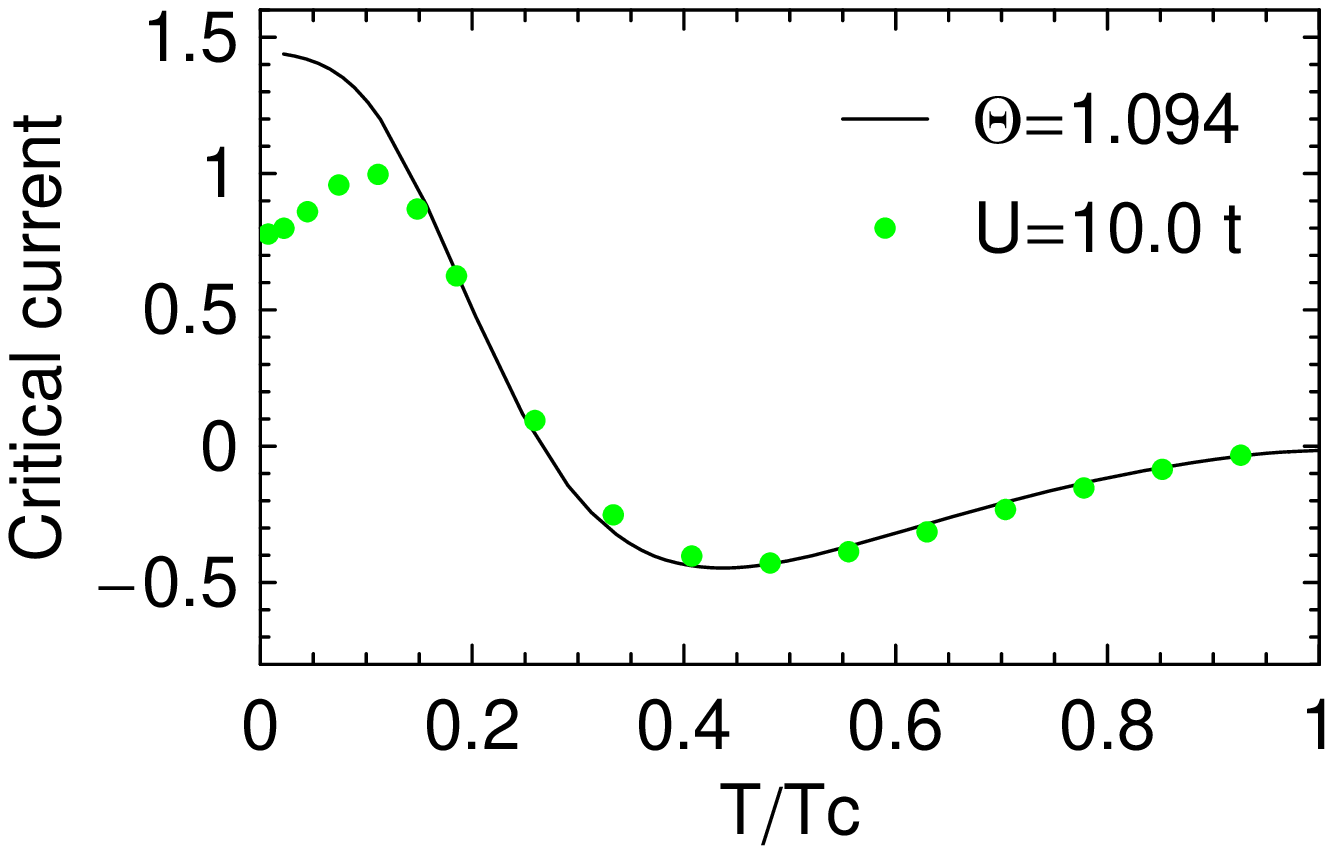}
\end{minipage}
\caption{(Color online) Critical current $J_c(T)$ vs temperature for
(left) d/AFeven/d and (right) d/AFodd/d junctions. The green dots
display the same $U=10t$ BdG data points shown in Fig.
\ref{fig:1}a,b (normalized to 1.0). The solid curves are
quasiclassical fits where the special choice $\cos k_y/\sqrt{2}=0.7$
has been taken, and the transparency $D$ has been adjusted to fit
the BdG data.} \label{fig:25}
\end{figure}

Above we have extracted the form of the current characteristics in
d/AF/d junctions via both numerical BdG methods and an analytical
quasiclassical approach. An obvious question is how well these two
methods agree. To this end, in Fig. \ref{fig:25} we plot again the
BdG results (normalized to 1.0) for the case $U=10t$ with $L=6$
(left) and $L=5$ (right), and show self-consistent quasiclassical
fits to these curves. Here $\sin \Theta = m/2t\cos
k_y/\sqrt{2}/[1+(m/4t\cos k_y/\sqrt{2})^2]$ and the special choice
$\cos k_y/\sqrt{2}=0.7$ has been taken. The transparency $D$ has
been adjusted to fit the BdG data. As seen, there is overall very
good agreement. At low $T$ some discrepancy can be detected, which
we believe originates from the finite interlayer thickness used in
the BdG calculations and/or the different bandstructures (circular
vs square Fermi surface in the quasiclassical and BdG approach,
respectively). Disregarding any explicit $k_y$ dependence of the
transparency coefficients and the $\Theta$ parameter in the
quasiclassical calculations may also play a role.

Experimental detection of $0-\pi$ transitions in $d$/AF/$d$
junctions may be possible in sandwich structures of high-doped and
un-doped high-$T_c$ materials similar to what was constructed for
c-axis junctions\cite{bozovic}. Recently, Oh {\it et
al.}\cite{oh:2005}, invented a spatially controlled doping method
and fabricated in-plane 100 $d$/AF/$d$ junctions. Our results show
that the fabrication of similar 110 junctions and a detailed study
of their Josephson current characteristics holds the promise of
several unusual properties as shown in Figs.
\ref{fig:1}-\ref{fig:2}. Realistic junctions will contain regions
with varying interlayer thickness, but if these are sufficiently
few, the regions with shortest thickness should dominate the
current. Alternatively, one needs to average the current over
interface imperfections. $J_c$ in even junctions dominates at low
$T$ only in the limit of large $U$. Then we have a small $\Theta$
and 0-junction with a low-$T$ anomaly in $J_c$. Otherwise critical
currents in even and odd junctions are of the same order. For
$\Theta>\pi/2$ (i.e. $m<4t$) the currents have identical signs at
all $T$ (0-junctions). For $\Theta<\pi/2$, the $\pi$-junction state
arises in odd junctions near $T_c$, resulting in an overall
cancellation of odd and even contributions to the current.

\section{Grain boundary junctions}
\begin{figure}[t]
\includegraphics[width=8.5cm,height=7.0cm]{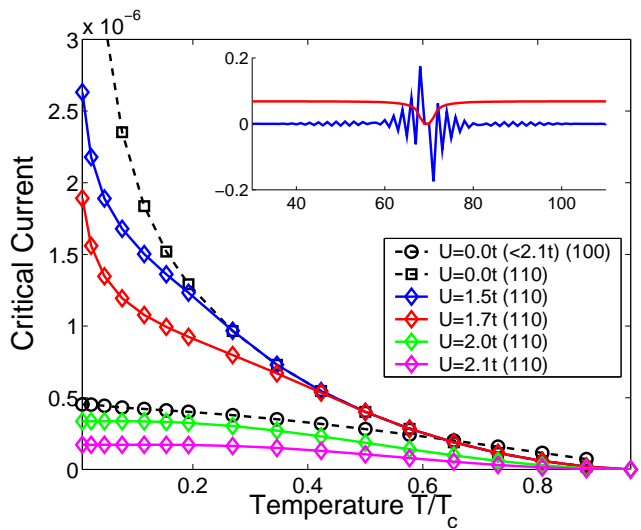}
\caption{(Color online). $J_c(T)$ versus $T$ for 110 and 100
$d$/I/$d$ junctions with AF correlations in the $d$ leads which have
doping level $x=0.1$. The I region is modeled with $L=2$ and a
potential of $V=30t$. In the 100 case, the same curve (black
circles) is obtained for all $U\leq 2.1t$. (inset) Example of
surface-induced magnetization (blue) and the suppression of
$\Delta^d$ (red) at a 110 interface, shown here for $U=2.0t$.
\label{fig:3}}
\end{figure}

Finally, we turn to the question of $J_c$ through grain boundaries,
where a strong discrepancy between theory and experiment has been
known for some time: when the GB is modeled as a $d$/I/$d$ junction
the zero-energy state existing in the 110 orientation results in a
large low $T$ increase of $J_c$ as compared to the 100 orientation
(see dashed lines in Fig. \ref{fig:3}). However, the opposite
behavior is obtained in experiments: $J_c$ is largest for 100
orientations and drops exponentially with increased angle between
the GB and the crystal axis\cite{hilgenkamp02}. We model the GB
using Eq.\eqref{hamiltonian} in a $d$/I/$d$ geometry with a
potential $V(n_{i\uparrow}+n_{i\downarrow})$ inside the insulating
layer (I) and $U \neq 0$ in the leads only. For sufficiently small
$U$, magnetization is absent in the superconducting leads, but the
magnetic correlations can lead to instabilities near interfaces that
suppress the dSC order parameter\cite{Ohashi99,honerkamp:2000}, as
shown in the inset of Fig. \ref{fig:3}. The main body of Fig.
\ref{fig:3} shows $J_c(T)$ for a range of $U$ all small enough not
to induce magnetization in the bulk. Contrary to the 100
orientation, $J_c$ through 110 GB can be significantly reduced by
surface-induced magnetic order for $T<T_M$, where $T_M$ is the
critical temperature for the surface magnetization. In fact, as seen
from Fig. \ref{fig:3} there exists a range of $U$ where $J_c$ at low
$T$ becomes smaller in the 110 orientation compared to the 100. This
shows the importance of competing surface effects even though a
complete understanding of the physics of GB junctions requires more
detailed microscopic calculations.

\section{Conclusions}

We have studied the dc Josephson current through d/AF/d tunnel
junctions as a function of interlayer thickness and temperature
using both numerical BdG diagonalization and analytical
quasiclassical methods. For an odd (even) number of
antiferromagnetic chains in the interlayer, the current
characteristics of 110 oriented interfaces display
$\pi$($0$)-junction behavior. In addition d/AFodd/d junctions can
exhibit $\pi-0$ transitions as a function of temperature. We have
shown that in terms of the spin-mixing parameter $\Theta$, the
condition for the latter is given by $\Theta<\frac{\pi}{2}$. This is
the opposite regime as compared to leads with $s$-wave pairing
symmetry where temperature-induced $\pi-0$ transitions take place
for $\Theta>\frac{\pi}{2}$. Another important difference between
s/AF/s and d/AF/d junctions exists for the 100 orientation, where
$d$-wave junctions always are 0-junctions whereas this is not the
case for $s$-wave superconductors. Finally we studied grain boundary
junctions modeled as d/I/d junctions but with subdominant magnetic
correlations in the superconducting leads allowing for
interface-induced magnetism near grains which tend to suppress the
$d$-wave order parameter. We showed that this mechanism can lead to
larger critical currents for the 100 orientation than for 110, in
qualitative agreement with experiments.

\section{Acknowledgements}

Partial support for this research was provided by DOE Grant
DE-FG02-05ER46236. Yu. S. B. acknowledges the support of RFBR grant
05-02-17175. Numerical calculations were performed at the University
of Florida High-Performance Computing Center (http://hpc.ufl.edu).

\end{document}